\documentstyle[aps,epsf]{revtex}
\begin{document}
\title{{\bf On  impact parameter dependence
 of low-$x$ structure functions}}
\author{
S. M. Troshin, N. E. Tyurin\\
\small \it
 Institute for High Energy Physics,\\
\small \it
 Protvino, 142284 Russia}
\normalsize

\maketitle
\begin{abstract}
We consider impact parameter dependence of the
polarized and unpolarized structure functions.
 Unitarity does
not allow factorization  of the structure functions
over the Bjorken $x$ and the impact parameter $b$ variables.
 On the basis of the
particular geometrical model approach we conclude
that  spin of constituent
quark may have a significant orbital angular momentum
component which can manifest itself through the peripherality
of the spin dependent structure functions.
\end{abstract}

\section*{Introduction}
The behaviour and dependence of the  structure functions
on the Bjorken $x$ is among the most actively discussed
subjects in the unpolarized and polarized
deep-inelastic scattering. The particular role here
belongs to the small $x$ region where asymptotical properties
of the strong interactions can be studied. The characteristic
 property of the low-$x$ region is an essential contribution of
nonperturbative effects \cite{bj,pasc} and one of the
possible ways to treat this region is the construction and
application of models.
Of course, the shortcomings of  any model approach
to the study of this nonperturbative region is  evident.
However, one could hope to gain from these models an information
which cannot  be obtained by perturbative methods
(cf. \cite{bj}).
Among  possible extensions in these studies are the
  considerations
of the geometrical features of the structure functions, i.e.
the dependence  of the structure functions on the transverse
coordinates or the impact parameter. This subject is not
new. The importance of
the parton distributions in the transverse plane was
stressed in \cite{bj} and, e.g. a brief model discussion
was recently given in \cite{tts}. This work is
a revised and extended version of the latter one.
  As it has been
demonstrated  \cite{burk} the $b$-dependent parton distribution
can be related to the Fourier transform of the off-forward matrix
elements of parton correlation functions in the limiting case
of zero skewedness.
 Impact parameter  dependence  would allow one to gain an
information on the spatial distribution of the partons inside
the parent hadron and the spin properties of the nonperturbative
intrinsic hadron structure. The geometrical properties of
structure functions  play an important role
 under analysis of
the lepton--nuclei deep--inelastic scattering and in the
hard production in the heavy--ion collisions.

\section{Interpretation of \lowercase {$b$}--dependent
structure functions at small \lowercase {$x$}}

  We suppose that the
deep--inelastic scattering is determined by the aligned-jet
mechanism \cite{bj}  and consider the $b$--dependence of
 the structure functions
along the lines used in \cite{usdif}.  The aligned-jet mechanism
  is  an essentially nonperturbative
and allows one to relate structure functions with the
discontinuities of the amplitudes
 of quark--hadron elastic scattering.
These relations are the following \cite{jj,sf}
 \begin{equation}\label{def1}
q(x)  =  \frac{1}{2}\mbox{Im}[F_1(s,t)+F_3(s,t)]|_{t=0},\quad
\Delta q(x)  =  \frac{1}{2}\mbox{Im}[F_3(s,t)-F_1(s,t)]|_{t=0},
\quad \delta q(x)  =  \frac{1}{2}\mbox{Im} F_2(s,t)|_{t=0}.
\end{equation}
 The functions $F_i$ are  helicity amplitudes
for the elastic quark-hadron scattering in the standard
notations for the nucleon--nucleon scattering.
We consider high energy limit or the region of small
$x$.

The structure functions obtained according to the above
formulas should be multiplied by the factor $\sim 1/Q^2$
-- probability that such aligned--jet configuration
 occurs \cite{bj}.

The amplitudes $F_i(s,t)$ are the corresponding Fourier-Bessel
transforms of the functions $F_i(s,b)$.

The relations Eqs. (\ref{def1}) will be used as a starting
point under definition of the structure functions which
depend on impact parameter. According to these relations it
is natural to give the following operational  definition:
\begin{equation}\label{def2}
q(x,b)  \equiv  \frac{1}{2}\mbox{Im}[F_1(x,b)+F_3(x,b)],\quad
\Delta q(x,b)  \equiv
\frac{1}{2}\mbox{Im}[F_3(x,b)-F_1(x,b)],\quad
\delta q(x,b)  \equiv  \frac{1}{2}\mbox{Im}F_2(x,b),
\end{equation}
and $q(x)$, $\Delta q(x)$ and $\delta q(x)$ are the integrals
over $b$ of the corresponding $b$-dependent distributions, i.e.
\begin{equation}\label{int1}
q(x)=\frac{Q^2}{\pi^2 x}\int_0^\infty bdb q(x,b),\quad
\Delta q(x)=\frac{Q^2}{\pi^2 x}\int_0^\infty bdb \Delta q(x,b),
\quad
\delta q(x)=\frac{Q^2}{\pi^2 x}\int_0^\infty bdb \delta q(x,b).
\end{equation}

The functions $q(x,b)$, $\Delta q(x,b)$ and $\delta q(x,b)$ depend
also on the variable $Q^2$  and have  simple  interpretations, e.g.
the function $q(x,b,Q^2)$ represent probability to find in the
hadron  a quark $q$ with  fraction of its longitudinal momenta
$x$ at the transverse distance
\[
b\pm \Delta b,\quad
 \Delta b\sim 1/Q
\]
 from the hadron geometrical center.
 Interpretation of the spin distributions directly follows
from their definitions:  they are the differences of the probabilities to find
quarks in the two spin states with longitudinal or transverse
directions of the quark and hadron spins.

It should be noted that the unitarity plays crucial role
in the direct probabilistic
interpretation of the function $q(x,b)$. Indeed due to unitarity
$0\leq q(x,b)\leq 1$.
The integral $q(x)$ is a quark number density which
is not limited by unity and can have arbitrary non--negative
value.
Thus, the given definition of the $b$--dependent structure
functions is self-consistent.  Of course,
spin distributions $\Delta q(x,b)$ and $\delta q(x,b)$ are not positively
defined.

\section{Unitarity and \lowercase {$b$}--dependence of structure functions}

The unitarity can be fulfilled through the $U$--matrix representation
for the helicity amplitudes of elastic quark--hadron scattering.
In the impact parameter representation the expressions for the
helicity amplitudes are the following
\begin{equation}
F_{1,3}(x,b)  = { U_{1,3}(x,b)}/{[1-iU_{1,3}(x,b)]},\quad
F_2(x,b)  =  {U_2(x,b)}/{[1-iU_1(x,b)]^2}\label{fu}
\end{equation}
Unitarity requires Im$U_{1,3}(x,b)\geq 0$. The $U$--matrix
 form of unitary
representation contrary to the eikonal one does not generate
itself essential singularity in the complex $x$ plane
 at $x\to 0$ and   implementation of unitarity can be performed
easily.

The model which provides explicit form of helicity functions
$U_i(x,b)$ has been described elsewhere \cite{usdif}.
A hadron consists of the constituent quarks aligned in the
longitudinal direction and embedded into the nonperturbative
vacuum (condensate). The constituent quark appears
as a quasiparticle, i.e. as current valence quark surrounded by
the cloud of quark-antiquark pairs of different flavors.
 We refer to effective QCD approach and  use the NJL model
\cite{njl} as a basis. The  Lagrangian  in addition to the
four--fermion interaction ${\cal L}_4$ of the original NJL model includes
the six--fermion $U(1)_A$--breaking term
${\cal{L}}_6\propto K(\bar u u)(\bar d d)(\bar s s)$ \cite{hoof}.
Transition to partonic picture
 is described by the introduction of a momentum cutoff
 $\Lambda=\Lambda_\chi\simeq 1$ GeV, which corresponds to the scale
of chiral symmetry spontaneous breaking \cite{jaffe}.

This  picture for a hadron structure implies  that  overlapping  and
 interaction of peripheral condensates in hadron  collision  occurs
 at the first stage. In the overlapping region the condensates
 interact and as a result virtual massive quark pairs appear.  Being
 released a part  of  hadron  energy  carried  by  the peripheral
 condensates goes to generation of massive quarks. In another words
 nonlinear field couplings  transform  kinetic  energy  into  internal
 energy  of dressed quarks. Of course,
 number of such  quarks fluctuates.  The average number of quarks in
 the cosidered case is
 proportional  to convolution  of the condensate distributions
  $D^{Q,H}_c$
 of the colliding constituent quark and hadron:
 \begin{equation} N(s,b) \simeq N(s)\cdot D^Q_c
 \otimes D^H_c, \end{equation} where the function  $N(s)$  is
 determined  by  a transformation thermodynamics    of  kinetic
 energy  of interacting  condenstates to the internal energy of
 massive quarks. To estimate the $N(s)$ it is feasible to assume that
 it is proportional to the  maximal possible energy dependence
 \begin{equation} N(s) \simeq \kappa  {(1-\langle x_Q\rangle )
 \sqrt{s}}/{\langle m_Q\rangle}, \label{kp} \end{equation} where
  $\langle x_Q \rangle
 $ is the average fraction of energy carried by the constituent quarks,
 $\langle m_Q\rangle $
 is the mass scale of constituent quarks.
In the model each of the constituent valence quarks located in the
 central part of the hadron
 is supposed to scatter in a quasi-inde\-pen\-dent way by the
 produced virtual  quark pairs at given impact parameter and by  the
  other valence quarks.
When smeared over longitudinal momenta
the  scattering  amplitude  of constituent valence quark $Q$ may
 be represented in
 the form \begin{equation} \langle f_Q (s,b) \rangle = [N(s,b) +
 N-1]\langle V_Q(b) \rangle,\label{sme} \end{equation} where $N=N_H+1$
 is  the
 total  number  of  quarks  in the system of the colliding  constituent quark
 and hadron  and $\langle
 V_Q(b) \rangle$ is  the smeared amplitude  of single quark-quark
 scattering.
In this approach the elastic scattering amplitude satisfies  the
 unitarity  since it is constructed as a solution  of  the
 following equation \begin{equation} F = U + iUDF
 \label{xx} \end{equation} which is presented here in operator form.
  The function $U(s,b)$  (generalized  reaction matrix)
  --- the basic dynamical quantity of  this approach ---
  is then
 chosen as a product of the averaged quark amplitudes \begin{equation}
 U(s,b) = \prod^{N}_{Q=1} \langle f_Q(s,b)\rangle \end{equation} in
 accordance  with assumed quasi-independent  nature  of  valence
 quark scattering.
 The strong
interaction radius of the constituent quark $Q$ is determined
by its Compton wavelength and
the $b$--dependence of the function $\langle f_Q \rangle$ related to
  the quark formfactor $F_Q(q)$ has a simple form $\langle
 f_Q\rangle\propto\exp(-m_Qb/\xi )$.
The helicity flip transition, i.e. $Q_+\rightarrow Q_-$
 occurs when the valence   quark
knocks out  a quark with the opposite  helicity
and the same  flavor \cite{abk}.

The explicit expressions for the helicity functions $U_i(x,b)$
at small $x$ have been obtained from the  functions
$U_i(s,b)$ \cite{usdif} by the substitute $s\simeq Q^2/x$
and at small values of $x$ they are the following:
\begin{equation}
U_{1,3}(x,b)=U_0(x,b)[1+\beta _{1,3}(Q^2)m_Q \sqrt{x}/Q],\quad
U_2(x,b)=g_f^2(Q^2)\frac{m_Q^2x}{Q^2}\exp[-2(\alpha -1)m_Qb/\xi]U_0(x,b),
\label{uis}
\end{equation}
where
\begin{equation}\label{u0}
U_0(x,b)=i\tilde U_0(x,b)
=i\left[\frac{a(Q^2)Q}{m_Q\sqrt{x}}\right]^{N}\exp[-Mb/\xi].
\end{equation}
  $a$, $\alpha$, $\beta$, $g_f$ and $\xi$ are the model
parameters, some of them in this particular case of quark-hadron
scattering depend on the virtuality $Q^2$. The
 meaning of these parameters  is not
crucial here; note only that $m_Q$ is
the average mass of constituent quarks in the quark-hadron system
of $N=N_H+1$ quarks and $M$ is their total mass, i.e.
$M=\sum_{i=1}^{N}m_i$. We  consider here for simplicity
 pure imaginary case.
We need to keep the subleading terms in the expressions for
$U_1(x,b)$ and $U_3(x,b)$ since the $\Delta q(x,b)$
is determined by their difference.
For  $U_2(x,b)$ one can keep only leading term.

 Then using Eqs. (\ref{fu}) we obtain at small
$x$:
\begin{equation} \label{qxb}
q(x,b)  = \frac{\tilde U_0(x,b)}{1+\tilde
U_0(x,b)},\quad
\Delta q(x,b) =  \frac{\beta_-(Q^2) m_Q\sqrt{x}}{Q}\frac{\tilde U_0(x,b)}
{[1+\tilde U_0(x,b)]^2},
\end{equation}
\begin{equation} \label{dqxb}
\delta q(x,b) =  \frac{g_f^2(Q^2)m_Q^2x}{Q^2}
\exp[-{ 2(\alpha -1)m_Qb}/{\xi}]
\frac{\tilde U_0(x,b)}{[1+\tilde U_0(x,b)]^2},
\end{equation}
where $\beta_-(Q^2)=\beta_3(Q^2)-\beta_1(Q^2)$.
From the above expressions it follows that $q(x,b)$ has a central
$b$--dependence, while $\Delta q(x,b)$ and $\delta q(x,b)$ have
 peripheral profiles. Their qualitative dependence on the impact
  parameter
$b$ is depicted in Fig. 1. The function $\Delta q(x,b)$ has a
  maximum located at
\[
b_{max}(x)=\frac{\xi N}{M}\ln[\frac{a(Q^2)Q}{m_Q\sqrt{x}}].
\]
\begin{figure}[hbt]
 \hspace*{5cm}
 \epsfxsize=80  mm  \epsfbox{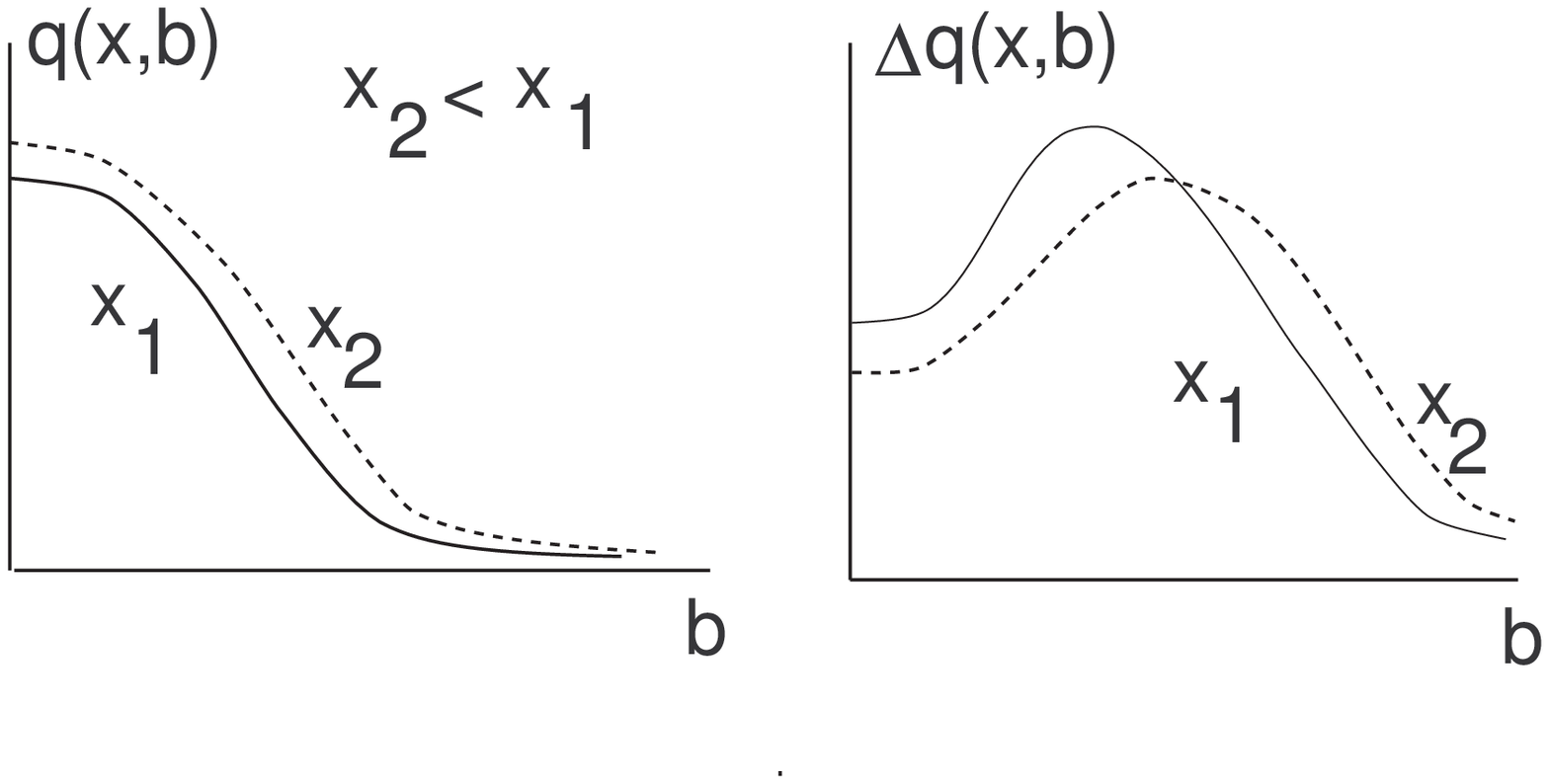}
\caption[junk]{{\it
$b$--dependence  of the  structure functions
 $q(x,b)$ and $\Delta q(x,b)$ at low-$x$}}.
 \end{figure}

From Eqs.(\ref{qxb},\ref{dqxb}) it follows that factorization
of $x$ and $b$ dependencies is not allowed by unitarity and this
provides constraints for the model parameterizations of structure
functions which depend on $x$ and $b$ variables. Indeed, it is clear
 from Eqs.(\ref{qxb},\ref{dqxb}) that factorized form of the input
  ``amplitude'' $\tilde U_0(x,b)$ cannot survive after unitarization
   due to the presence of the denominators.
It is to be noted here that from the relation
of impact parameter distributions with the off-forward parton
distributions  \cite{burk}
it follows that the same conclusion on the absence of
factorization is also valid for the off-forward parton distributions
with zero skewedness.

The following relation between the structure functions $\Delta q(x,b)$
and $\delta q(x,b)$ can also be inferred from the above model-based
 formulas
\begin{equation}\label{rel}
\delta q(x,b)=c(Q^2)\frac{\sqrt{x}}{Q}\exp(-\gamma b)\Delta q(x,b).
\end{equation}
Thus, the function $\delta q(x,b)$ which describes  transverse
spin distribution is suppressed by the factors
$\sqrt{x}$ and $\exp(-\gamma b)$, i.e. it has a more central
profile. This suppression also reduces double-spin transverse
asymmetries in the central region in the Drell-Yan production
compared to the corresponding longitudinal asymmetries.

The strange quark structure functions have also a more central
$b$--dependence than in the case of $u$ and $d$ quarks.
 The radius of the
corresponding quark matter distribution follows from Eq. (\ref{dqxb})
and is the following
\begin{equation}\label{rad}
R_q(x)\simeq \frac{1}{M}\ln{Q^2}/{x}
\end{equation}
 while the ratio of the strange quark distributions to the
light quark distributions radii is given by the corresponding
constituent quark masses,
i.e. for the nucleon this ratio would be
\begin{equation}\label{ratr}
R_s(x)/R_q(x)\simeq (1+\frac{\Delta m}{4m_Q})^{-1},
\end{equation}
where $\Delta m = m_S-m_Q$.

Time reversal invariance of strong interactions allows
one to
write down relations similar to Eqs.(\ref{def1}) for the
fragmentation functions also and obtain expressions
for the fragmentation functions $D_q^h(z,b)$, $\Delta D_q^h(z,b)$,
$\delta D_q^h(z,b)$ which have just the same dependence
on the impact parameter $b$
as the corresponding structure functions. The fragmentation
function $D_q^h(z,b,Q^2)$ is the probability for fragmentation
of  quark $q$ at  transverse distance
 $b ~ \pm ~ \Delta~ b$ ($\Delta b\sim 1/Q$)
into a hadron $h$ which carry the fraction
$z$ of the quark momentum. In this case $b$ is a transverse
distance between quark $q$ and the center of the
hadron $h$. It is positively
defined and due to unitarity obey to the inequality
$0\leq D_q^h(z,b)\leq 1$.
The physical interpretations of  spin--dependent
 fragmentation functions $\Delta D_q^h(x,b)$ and
$\delta D_q^h(x,b)$ is similar to that of  corresponding
spin structure function.
\section*{Discussions and Conclusion}

 It is interesting to note that the spin structure functions
have a peripheral dependence on the impact parameter contrary
to  central profile of the unpolarized structure function.
It could be related to the orbital angular momentum of quarks
inside the constituent quark.
The important point  what the origin of this orbital angular momenta
 is.
It was proposed \cite{spcon} to use an analogy with
 an  anisotropic extension of the theory of superconductivity
 which seems to match well with the  picture for a constituent
  quark.  The studies \cite{anders} of that theory show that the
   presence of anisotropy leads to axial symmetry of pairing
  correlations around the anisotropy direction $\hat{\vec{l}}$ and to
 the particle currents induced by the pairing correlations.  In
 another words it means that a particle of the condensed fluid
   is surrounded
 by a cloud of correlated particles  which rotate around
it with the
axis of rotation $\hat {\vec l}$.
 Calculation of the
orbital momentum  shows that it is proportional to the density of the
correlated particles.
Thus, it is clear that there is a direct analogy
between  this picture and that describing the constituent quark. An
axis of anisotropy $\hat {\vec l}$ can be associated with the
polarization vector of current valence quark located at the origin of the
constituent quark.  The orbital angular momentum $\vec L$ lies
along $\hat {\vec l}$.

Spin of constituent quark, e.g. $U$-quark, in the model used
 is given
by the  sum:
\begin{equation}\label{bal}
J_U=1/2=S_{u_v}+S_{\{\bar q q\}}+L_{\{\bar q q\}}=
1/2+S_{\{\bar q q\}}+L_{\{\bar q q\}}.
\end{equation}
In principle, $S_{\{\bar q q\}}$ and $L_{\{\bar q q\}}$
 can include contribution of
gluon angular momentum, however,
since we consider effective Lagrangian approach where gluon
 degrees of freedom are overintegrated, we do not concern
  problems of the
 separation and mixing of the quark  angular momentum and
 gluon effects in QCD (cf.  \cite{kiss}).
  In the NJL--model
\cite{jaffe} the six-quark
 fermion operator simulates the effect of gluon operator
$\frac{\alpha_s}{2\pi}G^a_{\mu\nu}\tilde G^{\mu\nu}_a$,
where $G_{\mu\nu}$ is
the gluon field tensor in QCD.
It is worth to note here that in general large gluon orbital
 angular momentum
is expected to be almost canceled by gluon spin
 contribution \cite{rat}.

The value of the orbital
 momentum contribution into the spin of constituent quark can be
 estimated according to  the relation between
 contributions of current quarks into a proton spin and corresponding
contributions of current quarks into a spin of the constituent quarks
and that of the constituent quarks into  proton spin \cite{altar}:
\begin{equation} (\Delta\Sigma)_p = (\Delta U+\Delta
D) (\Delta\Sigma)_U,\label{qsp} \end{equation}
where $(\Delta\Sigma)_U=S_{u_v}+S_{\{\bar q q\}}$.
The value of $(\Delta\Sigma)_p$ was measured in the deep--inelastic
scattering.
Thus, on the grounds of the experimental data for polarized DIS
we arrive to conclusion that the significant part of the spin
of constituent quark in the model should be associated with
 the orbital angular momentum
of the current quarks inside the constituent one \cite{spcon}.

Then the peripherality of the spin structure functions
 can be correlated with the large contribution of the
 orbital angular momentum,
   i.e. with the  quarks
  coherent rotation.
Indeed, there is a compensation between the total spin
of the quark-antiquark cloud and its
  orbital angular momenta, i.e.
$L_{\{\bar q q\}}=-S_{\{\bar q q\}}$ and therefore
   this correlation follows
 if the above compensation has a local nature
 and valid for a fixed impact parameter.

The important role of orbital
 angular momentum was known
 long before EMC discovery \cite{sig} and reappeared after
 as  one of the  transparent explanations of the polarized
  deep-inelastic scattering data \cite{ell}. Lattice QCD calculations
  in the quenched approximation also indicate significant quark orbital
  angular momentum contribution to  spin of a nucleon \cite{math}.
  It is interesting
to find out possible experimental signatures of the peripheral
geometrical profiles of the spin structure functions and
the significant role of the orbital angular  momentum.
One of such indications could be an observation of the
different spatial distributions of charge and magnetization
at Jefferson Lab \cite{bur}. It would also be important to
have a precise data for the strange formfactor.\\
\bf{Acknowledgements:} \rm
 This work was supported in part by the Russian Foundation for
Basic Research under Grant No. 99-02-17995.


\end{document}